\documentclass[prb,superscriptaddress,showpacs,floatfix,twocolumn]{revtex4}

\usepackage{amssymb}
\usepackage{amsmath}
\usepackage{graphicx}

\newcommand{\w}{{\omega}}
\renewcommand{\vec}[1]{{\bf #1}}

\begin{document}

\title{
Large thermomagnetic effects in weakly disordered Heisenberg chains}
\author{E. Shimshoni}
\address{Department of Physics, Bar-Ilan University, Ramat-Gan 52900, Israel}
\author{D. Rasch}
\address{Institute for Theoretical Physics, University of Cologne, 50937
Cologne, Germany}
\author{P. Jung}
\address{Institute for Theoretical Physics, University of Cologne, 50937
Cologne, Germany}
\author{A. V. Sologubenko}
\address{II. Physical Institute, University of Cologne, 50937 Cologne, Germany}
\author{A. Rosch}
\address{Institute for Theoretical Physics, University of Cologne, 50937
Cologne, Germany}


\begin{abstract}
  The interplay of different scattering mechanisms can lead to novel
  effects in transport.  We show theoretically that the interplay of weak impurity
  and Umklapp scattering in spin-$1/2$ chains leads to a pronounced dip in
  the magnetic field dependence of the thermal conductivity $\kappa$ at a
  magnetic field $B \sim T$. In sufficiently clean samples, the
  reduction of the magnetic contribution to heat transport can easily become larger than $50\%$ and the
  effect is predicted to exist even in samples with a large exchange
coupling, $J \gg B$, where the field-induced magnetization is
small. Qualitatively, our theory might explain dips at $B \sim T$
observed in recent heat transport measurements on
copper pyrazine dinitrate, but a fully quantitative description is
not possible within our model.
\end{abstract}

\date{\today}
\maketitle

Some of the most fascinating manifestations of quantum many-body
physics in one-dimension (1D) can be found in spin-$\frac{1}{2}$
chain systems \cite{review,giamarchi}. In particular, the spin-$\frac{1}{2}$
Heisenberg model with antiferromagnetic nearest neighbor exchange
interactions is one of the most extensively studied paradigms. It
provides a remarkably non--trivial example of an exactly solvable
model \cite{bethe}, allowing a detailed analysis of its thermodynamic
properties. While integrability is special to this
particular model, its low energy properties are generic: they do not
differ essentially from low energy properties of other (non
integrable) spin chain models, with more general finite range
interactions. In essence, these systems
support Fermionic elementary excitations -- spinons -- which carry
spin and no charge. Their kinetic energy and interactions are
dictated by the exchange couplings in the chain, and their Fermi
momentum can in principle be tuned by a magnetic field $B$, which
plays the role of a chemical potential.

Thermodynamic properties of low-dimensional spin systems and
especially of the Heisenberg model are generally very well understood
allowing for a quantitative description of a broad range of
experiments. In comparison, the heat and spin transport \cite{spin-ex,sologubenko} in spin-systems
is considerably more complicated. For example, the heat conductivity $\kappa$
of the one-dimensional Heisenberg model is infinite at
finite temperatures as the heat current operator is a conserved
quantity for this idealized model. In real materials, the effects of
disorder, phonon coupling and spin interactions not captured by the
integrable Heisenberg model render the conductivity
finite. Nevertheless, it remains often very
large \cite{almost} as long as disorder is weak.

\begin{figure}
\includegraphics[width=0.9 \linewidth,clip=true]{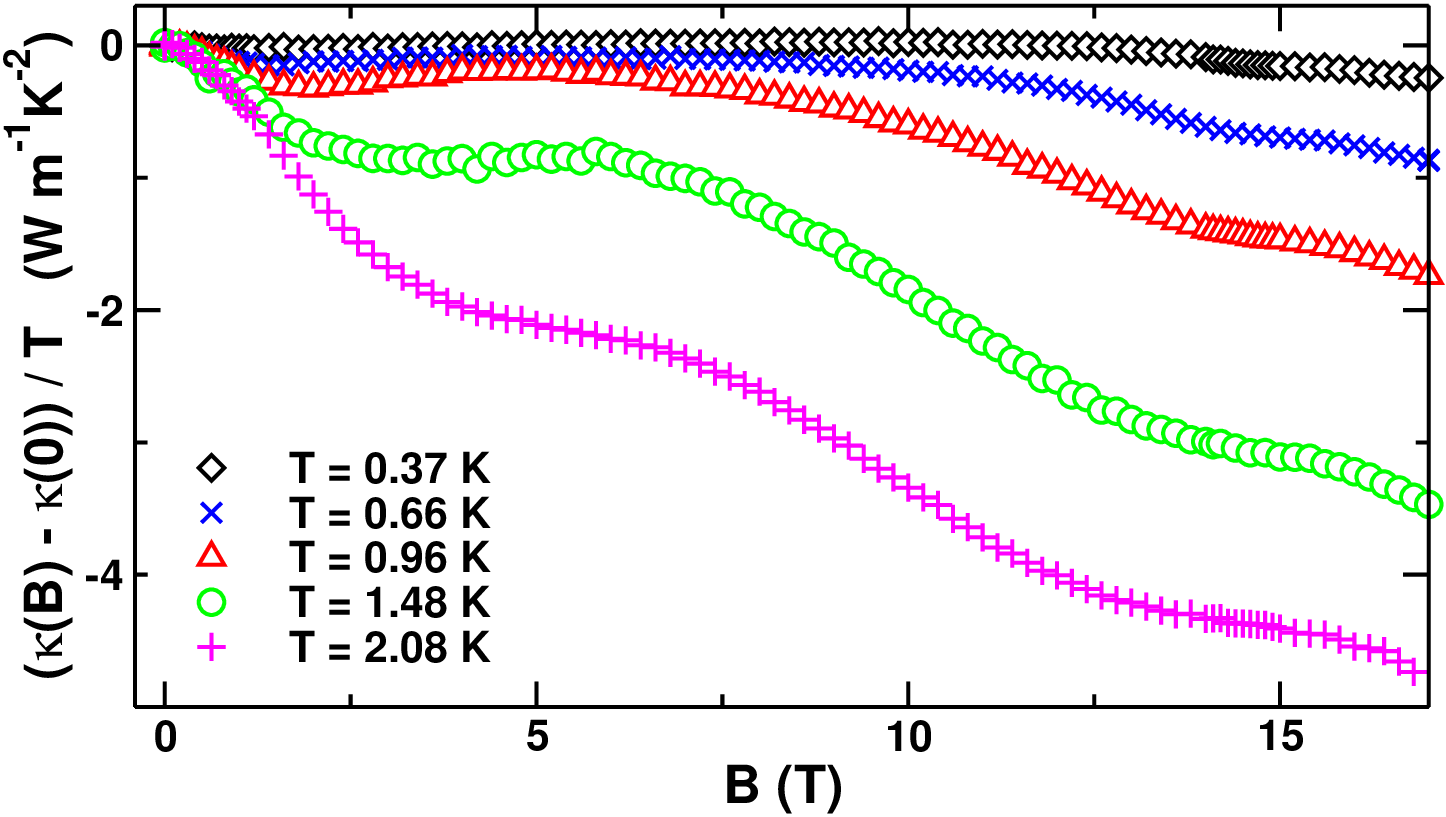}\\[3mm]
\includegraphics[width=0.9 \linewidth,clip=true]{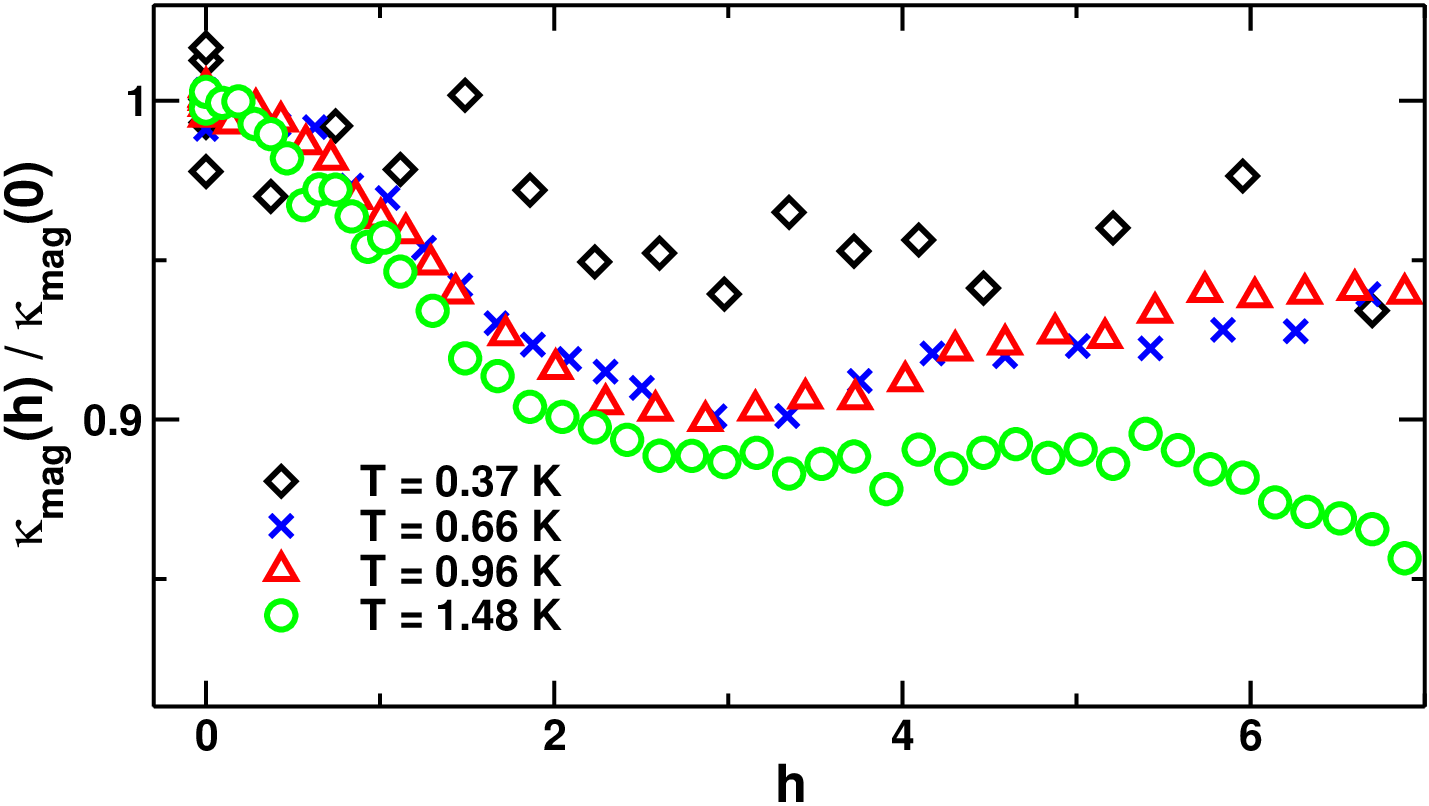}
\caption{(Color online). Upper figure: Field dependent thermal
conductivity $\kappa(B,T)-\kappa(0,T)$ in the spin-1/2 compound
CuPzN taken from Ref.~[\onlinecite{sologubenko}]. The downturn at
large fields arises as the spinon band is considerably depleted.
For $T=0$ the magnetization saturates at $B\approx 15$\,T. Lower
figure: The normalized magnetic part of the thermal conductivity,
$\kappa_{\rm mag}(B,T)/\kappa_{\rm mag}(0,T)$, as a function of
$h=\mu_B g B/(k_B T)$ shows a dip at $h\sim 3$. Note that there
might be a considerable error in the size of the dip as
$\kappa_{\rm mag}(B=0,T)$ could not be measured directly due to a
large phonon background, see text and
Ref.~[\onlinecite{sologubenko}] for details. \label{figExp}}
\end{figure}

In this paper we study how the strong interactions of the
Heisenberg model affect the heat transport in the system in the
presence of weak disorder beyond the well known effect that  disorder
is strongly renormalized by the interactions \cite{fisher}.  Our
study is directly motivated by recent experiments \cite{sologubenko} in the spin-1/2
chain compound copper pyrazine dinitrate
Cu(C$_4$H$_4$N$_2$)(NO$_3$)$_2$ (CuPzN) reproduced
in Fig.~\ref{figExp}.  In this system a small exchange coupling, $J/k_B
\approx 10.3$\,K, allows to polarize the system with moderate magnetic
fields $B$. While both the spins and the phonons contribute to the
heat transport at low temperatures $T$ with similar strength, the field
dependence can be used to separate the two effects.
The total heat conductivity $\kappa$ can be split
into a contribution arising purely
from phonons and a magnetic part,
$\kappa(B,T)=\kappa_{\rm ph}(T)+\kappa_{\rm mag}(T,B)$. Then, one can use the fact that
$\kappa_{\rm ph}(T)$ is practically independent of $B$ to extract
\begin{eqnarray}
\Delta
\kappa_{\rm mag}(T,B)&=&\kappa(T,B)-\kappa(T,B=0)\nonumber \\
&=&\kappa_{\rm mag}(T,B)-\kappa_{\rm mag}(T,B=0).
\end{eqnarray}
In CuPzN  $\Delta
\kappa_{\rm mag}(T,B)$ shows a pronounced dip as a function of magnetic
field for small $B$ and $T$. Interestingly, the position of this dip scales linearly with
the temperature, $g \mu_B B_{\rm min} \approx 3 k_B T$, pointing to a simple
underlying mechanism. A dip in $\kappa_{\rm mag}(B)$ has been observed in
numerical simulations \cite{zotos}
of classical spin-chains coupled to phonons but this dip occurs at $B
\sim J$ and does not scale with $T$.

Several processes can contribute to $\kappa_{\rm mag}$, the field
dependent part of $\kappa$. First, there is a positive
contribution from heat transported by the spin-chains. Second, the
heat conduction of the phonons is reduced when phonons scatter off
spin fluctuations.  Third, there is a contribution from spin-phonon
drag which is usually positive and can also become very large
\cite{ladder}. The positive sign of $\kappa_{\rm mag}$ in CuPzN suggests
that the first and possibly the third mechanism are dominating.

 Within a fermionic interpretation of the spin
excitations, the magnetic field enters as a chemical potential while
$T$ determines the broadening of the Fermi surface. Therefore a likely
interpretation of the experiment is that the characteristic dip arises
when the Fermi surface moves away from the momentum $k_F=\pi/2a$,
corresponding to a half-filled system with lattice spacing $a$. The commensurate filling for
$B=0$ is very special as it allows for Umklapp scattering: two
spinon excitations can be transferred from the left to the right Fermi
surface and the excess momentum $4 k_F = 2 \pi/a$ can be absorbed by the
underlying lattice.
As this scattering mechanism is only effective in the vicinity of the commensurate point,
it is exponentially suppressed for $g \mu_B B >
k_B T$. We will show that besides the Umklapp scattering it is
necessary to include the effects of impurities to get structures at $B
\sim T$; in the absence of disorder, the presence of certain
approximate conservation laws \cite{shimshoni03} prohibits a relaxation
of the heat current by the leading Umklapp process [see (\ref{umklapp}) below] alone.

{\em Model and methods:} We investigate the one-dimensional Heisenberg model in the
presence of a magnetic field $B$ and weakly disordered exchange
couplings, $\delta J_i \ll J$
\begin{eqnarray}\label{H0}
H&=& -  \sum_i (J+\delta J_i) \,  \vec{S}_i \cdot \vec{S}_{i+1}- g \mu_B B \sum_i S_z.
\end{eqnarray}
In the following, we will be interested in the limit of small magnetic
fields and temperatures, $B,T \ll J$. More precisely, all calculations
are done only to leading order in $1/\ln[(T,B)/J]$. The interactions
lead to a strong renormalization of disorder \cite{fisher}, but we
assume that the temperatures are sufficiently high, such that
 the renormalized disorder remains weak, $\delta J \ll
\sqrt{J T}$.

For $B,T,\delta J_i \ll J$ one can use the powerful techniques of
bosonization to describe the low-energy properties of the system.
It is useful to split the effective low-energy Hamiltonian into
three pieces
\begin{eqnarray}\label{HLL}
H&=&H_{LL}+H_U+H_{dis} \\
 H_{LL}&=&v\int \frac{dx}{2 \pi} \left( K
(\partial_x \theta)^2+\frac{1}{K} (\partial_x \phi)^2 \right)\\
H_U &=&  \frac{ g}{(2 \pi a)^2} \int dx [e^{i \Delta k x} e^{ i 4
\phi }  + h.c.]\label{umklapp}\\
H_D&=&\frac{1}{2 \pi a}\int dx \,  \eta(x)[ i e^{ i 2 \phi } + h.c.] \label{dis}
\end{eqnarray}
where (using the notation of
[\onlinecite{giamarchi,shimshoni03}])
 $\partial_x \phi$ denotes fluctuations of the magnetization in the $z$ direction,
$\theta$ is the conjugate variable with
$[\phi(x),\partial_x \theta(x')]=i \pi \delta(x-x')$, and we use units
where $k_B=1, \hbar=1$. For the
spin-rotationally invariant Heisenberg model, the velocity and the
Luttinger parameters at the low-energy fixed point are given by
$v=\frac{\pi}{2} J a$ and $K=1/2$, respectively. The Umklapp term $H_U$
describes the scattering of spinons from one Fermi point to
the other. At a finite magnetization, the Fermi momentum
$k_F=\frac{\pi}{2 a} (1+2 \langle S_z \rangle)$ deviates from
$\pi/2a$ and therefore the excess momentum $\Delta k=4 k_F-\frac{2
\pi}{a}=4 \pi \langle S_z \rangle/a$ cannot be absorbed by the
crystalline lattice. For $\Delta k=0$ the Umklapp scattering is a
marginally irrelevant operator whose strength decreases
logarithmically with $T$ (see below) while for $v \Delta k \gg k_B T$
it is exponentially suppressed in a clean system. The effects of
disorder, or more precisely from components of the disorder
potential oscillating with momentum $2 k_F$, is described by
$H_D$ which models the scattering from one  Fermi point to the other with $\eta(x) \sim \delta J(x)$.
Here we assume uncorrelated disorder with $\langle \eta(x) \eta(x')\rangle =
D_{dis} \delta(x-x')$.

To calculate the heat conductivity we use the so-called memory
matrix formalism \cite{memory} as in [\onlinecite{shimshoni03}]. Within
this approach one calculates a matrix of relaxation rates for a
given set of modes. As has been discussed in detail in
[\onlinecite{bounds}], in general this method allows to calculate a lower
bound to the conductivity. The formalism gives precise results as
long as the relevant slow modes are included in the calculation.
For the present system, the essential step is to realize
\cite{roschPRL,shimshoni03} that in the absence of disorder the
operator $Q=J_H+v \frac{\Delta k}{4 K} J_s$ is conserved,
$[Q,H_{LL}+H_U]=0$. Here, $J_H=v^2 \int
dx\, \partial_x \theta \partial_x \phi$ is the heat current
associated with $H_{LL}$ and $J_s=v K  \int \partial \theta/\pi$ is the
spin-current. This can be seen by realizing that up to the prefactor $v^2$ the heat current
can  be identified with the momentum
operator, the generator of translations. The Umklapp term
describes a process where a momentum $\Delta k$ is generated and
therefore its commutator with $J_H$ is proportional to $v^2 \Delta
k$. Similarly, the spin current is changed by $-4 v$ as two
spinons with velocity $v$ are scattered into states with velocity
$-v$. Therefore the linear combination $Q=J_H+v \frac{\Delta k}{4 K}
J_s$ remains unaffected by Umklapp processes. In terms of the
original variables, $Q$ can be identified with
 $J^2 \sum_i \vec{S}_i (\vec{S}_{i+1} \times
\vec{S}_{i+2})$,  the heat current operator for $B=0$ which
commutes with the integrable Heisenberg model (\ref{H0})
in the absence of disorder, $\delta J_i=0$ (when longer range
interactions or interchain coupling break integrability, the lifetime
of $Q$ nevertheless remains exponentially large in a clean system, see
[\onlinecite{roschPRL,shimshoni03}]). We therefore set up the memory matrix
formalism in operator space spanned up by the two relevant currents
$J_H$ and $J_s$.

The decay rates of the currents are determined
\cite{memory,shimshoni03} from a $2 \times 2$ matrix $\hat M$ of
correlation functions (the `memory matrix'),
\begin{eqnarray}
M_{ij}=\lim_{\omega \to 0} \frac{{\rm Im} \,
\langle \partial_t J_i ; \partial_t J_j \rangle_{\w}}{\omega}
\end{eqnarray}
where $\langle A ; B \rangle_{\omega}$ denotes a retarded
correlation function of $A$ and $B$ evaluated at the real frequency
$\omega$ and $J_1=J_H$, $J_2=J_s$ are the two relevant
operators. Within the assumptions of our paper, we can treat both
Umklapp and disorder perturbatively and therefore it is sufficient to
evaluate all expectation values with respect to $H_{LL}$ with $K=1/2$
as the derivative $\partial_t J_i$ are already linear in the
perturbations $H_U$ and $H_{\rm dis}$.

The
heat conductivity per spin chain is obtained from
\begin{eqnarray}
\kappa_{\rm mag} \approx \frac{ \chi_{H}^2}{T} \left. \hat M^{-1}\right|_{11}=
\frac{ \chi_{H}^2}{T} \frac{M_{ss}}{M_{ss} M_{HH} -M_{sH}^2} \label{kappa}
\end{eqnarray}
where $\chi_H=\langle J_H;J_H\rangle_{\w=0} \approx \frac{\pi v
  T^2}{3}$ is the susceptibility of the heat current.

Separating the contributions from Umklapp and disorder,
$\hat{M}=\hat{M}_{U}+\hat M_{\rm dis}$, we find using straightforward
perturbation theory
\begin{eqnarray}
\hat{M}_U&=&\Gamma_U(B,T) \left(\begin{array}{cc}
\left(\frac{v \Delta k}{2}\right)^2& -\frac{v \Delta k}{2}\\[1mm]
-\frac{v \Delta k}{2}& 1  \end{array}\right) \label{mu}\\
\Gamma_U &=&  -\frac{g(T)^2 v (\Delta k)^2}{8 \pi^2} n_B'(v
  \Delta k/2)\label{gu}\\
\hat{M}_{\rm dis}&=&\frac{v \pi D_{\rm dis}}{8 a}  \left(\begin{array}{cc}
T & 0\\[1mm]
0 & \frac{2}{\pi^2 T}  \end{array}\right) \label{mdis}
\end{eqnarray}
where $n_B'(\omega)$ is the derivative of the Bose function, $n_B(\w)=1/(e^{\w/T}-1)$. The $T$ dependence
of $g(T)\sim \pi v/\ln(J/T)$ (see below) takes into account that the
Umklapp scattering is marginally irrelevant with respect to the clean
fixed point.

{\em Results:} The rather simple equations (\ref{kappa}-\ref{mdis}) describe the
complex interplay of disorder and Umklapp scattering. First, in the
absence of disorder, the heat conductivity is infinite \cite{footnote} as $\hat{M}_U$
has an eigenvalue $0$ reflecting the conservation of $Q$ described
above.
Second, for vanishing magnetic field, $\Delta k=0$, Umklapp scattering
plays no role and one obtains the well-known  result
$\kappa_{\rm mag}/T \sim T$. The mean free path decreases linearly in $T$ as the
disorder is strongly renormalized by the interactions \cite{fisher}. Third, for
finite magnetic field and $T \to 0$, Umklapp scattering
is exponentially suppressed, $\Gamma_U \sim e^{-v \Delta k/2 T}$,
as the Fermi energy has shifted away from commensurate filling and one
finds $\kappa_{\rm mag}(B \gg T)=\kappa_{\rm mag}(B=0)$. Here we have neglected the -- formally
subleading -- effect that the Luttinger liquid parameter $K$ and
therefore also the renormalization of the disorder potential
depend on $B$.

\begin{figure}
\includegraphics[width=0.9 \linewidth]{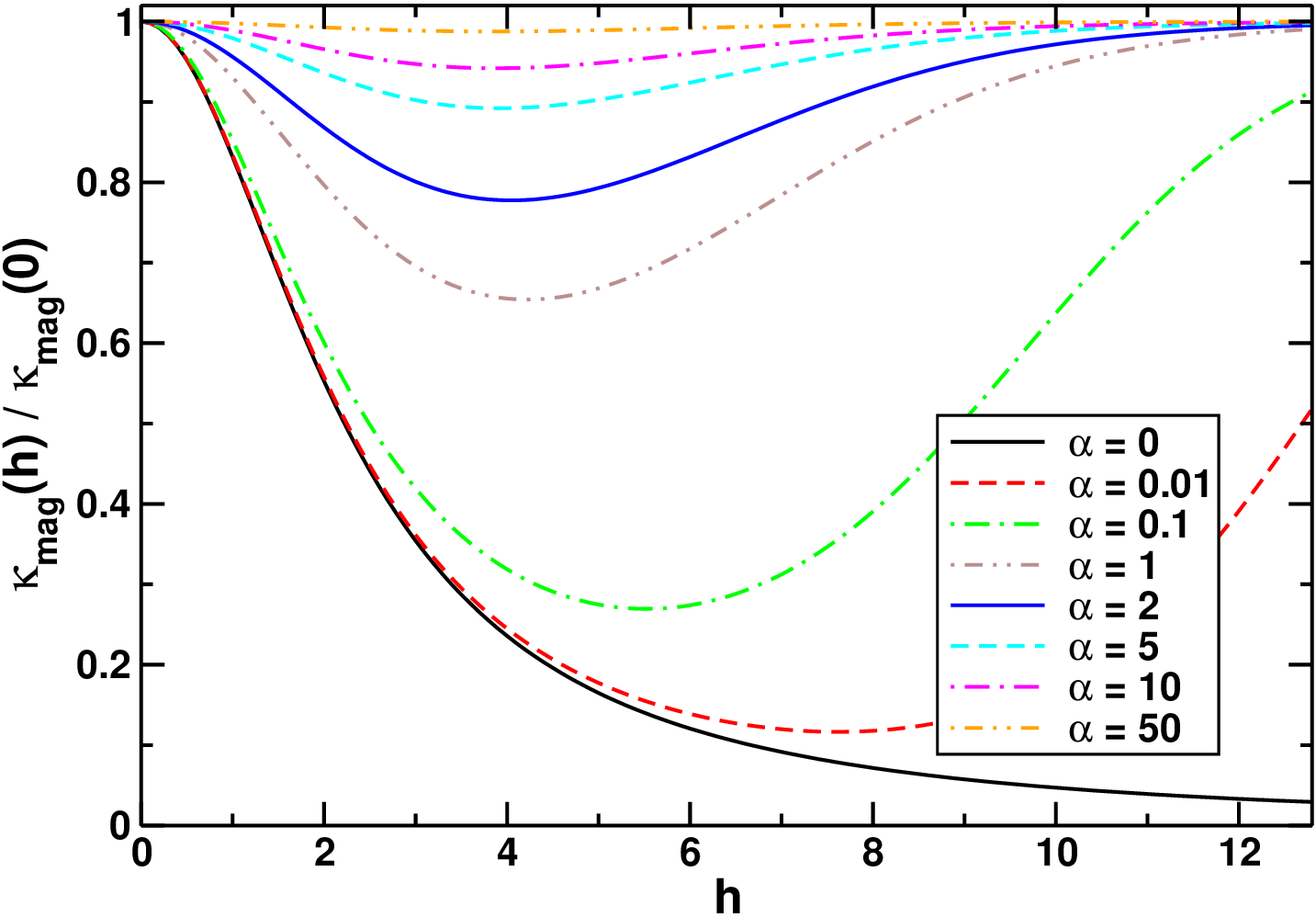}
\caption{ (Color online). Field dependence of the of the
normalized thermal conductivity as a function of the rescaled
field, $h=g \mu_B B/k_B T$. For moderate disorder $\alpha(T)$,
Eq.~(\ref{alpha}), $10 \gtrsim \alpha(T) \gtrsim 1$, a pronounced
dip in $\kappa_{\rm mag}(B)$ is predicted for $g \mu_B B\sim 4 k_B
T$. For even smaller values of $\alpha(T)$ the dip gets broader
and deeper, approximating
 the asymptotic behavior (\ref{kappaAs2}) (solid black line), up to larger and larger  fields. \label{figScaling} }\end{figure}

Upon increasing the magnetic field $B$,
 the Fermi surface is shifted and $\Delta k$
increases approximately linearly in $B$, $\Delta k \approx 4
\pi \chi B/a$ with
\cite{lukyanov} $\chi\approx g  \mu_B/(\pi^2 J)$. As argued above, for $\Delta k=0$ the Umklapp
scattering does not influence transport. Therefore, by raising $\Delta
k$  the effect of Umklapp scattering is switched on proportionally
to $(\Delta k)^2$. But upon increasing $\Delta k$ further, Umklapp
scattering is switched off for $v \Delta k \gg T$ or $B \gg T$. The
net result is a pronounced dip in $\kappa_{\rm mag}(B)$, see Fig.~\ref{figScaling}.

In the scaling limit of weak disorder and $1/\ln[(B,T)/J]\ll 1$ the
normalized conductivity $\kappa_{\rm mag}(B,T)/\kappa_{\rm mag}(B=0,T)$ is only a function
of the scaling variable $h=\mu_B g B/k_B T$ and a dimensionless variable
\begin{equation}
\alpha(T) = \frac{D_{\rm dis} v^2}{ (k_B T)^2 g(T)^2 a}
\approx \frac{8 k_B  v}{9 \pi  \tilde{g}(T)^2 \kappa_{\rm mag}(B=0,T)}\label{alpha}
\end{equation}
which parameterizes
the relative strength of (renormalized) disorder and Umklapp scattering ($\tilde{g}\sim 1/\ln[J/T]$ is defined below).
In these variables, we obtain
\begin{eqnarray}
\frac{\kappa_{\rm mag}(B,T)}{\kappa_{\rm mag}(0,T)}=
\frac{\pi^3 \alpha(T) - 2 \pi^2 h^2 n_B'(h)}{\pi^3 \alpha(T)-(2 \pi^2+4 h^2) h^2
  n_B'(h)}
\end{eqnarray}
with $n_B(h)=1/(e^{h}-1)$. As shown in Fig.~\ref{figScaling}, the
field dependence of the thermal conductivity is predicted to show a
pronounced dip at $B \sim T$. For small $\alpha(T) \ll h^2 n_B'(h)$,
i.e. weak disorder and not too strong fields, one finds
\begin{equation}
\frac{\kappa_{\rm mag}(B,T)}{\kappa_{\rm mag}(0,T)} \approx \frac{1}{1+2 h^2/\pi^2},\label{kappaAs2}
\end{equation} see
Fig.~\ref{figScaling}. This implies a strong reduction of $\kappa_{\rm mag}$ of
order $1$ for $\mu_B B \sim k_B T$ as long as the renormalized
disorder is sufficiently weak, $\alpha(T) \ll 1$.  This is the main result of the paper.
For large fields or
stronger effective disorder one obtains a small suppression of
$\kappa_{\rm mag}$,
\begin{equation}
\frac{\kappa_{\rm mag}(B,T)}{\kappa_{\rm mag}(0,T)}\approx 1-\frac{-4 n_B'(h) h^4}{\pi^3
  \alpha(T)},
  \end{equation} giving rise to a minimum at $h\approx 3.8$ of size
$0.63/\alpha(T)$.

To allow for a quantitative comparison to the experiment of
Ref.~[\onlinecite{sologubenko}], one needs to estimate $\alpha(T)$. For this, both
the strength of impurity scattering and the strength of the
renormalized Umklapp scattering $g(T)$ have to be
determined. Fortunately, for simple Heisenberg chains the latter is
known analytically from the Bethe Ansatz.  Translating our notations
to those used in [\onlinecite{lukyanov}], we obtain $g(T)=\tilde g(T) \pi^2 J a/2$ and $\tilde g$
is obtained by solving the equation \cite{lukyanov}
\begin{equation}
\frac{1}{\tilde g}+ \frac{\ln(\tilde g)}{2} = \ln\!\left( \frac{e^{1/4+\gamma} \sqrt{\pi/2} J}{T}\right)\label{g}
\end{equation} where
$\gamma=0.5772...$ is the Euler constant. Within the precision of our
calculation, $\tilde{g}\approx 1/\ln(J/T)$, but we use the more
precise formula (\ref{g}) which includes subleading corrections below. The disorder strength for the sample of CuPzN measured in
[\onlinecite{sologubenko}] can in principle be obtained from
$\kappa_{\rm mag}(B=0,T)$. Unfortunately, a large phonon background prohibits a
direct measurement of this quantity but a crude estimate, $\kappa_{\rm mag}
\approx 3.5\, T^2$\,Wm$^{-1}$K$^{-3}$, can be obtained from the
behavior of $\kappa$ at large fields (see [\onlinecite{sologubenko}] for
details). Using this estimate, we find for the heat conductivity per
spin chain $\kappa_{\rm mag} \approx 2.1\,10^{-18}\, T^2$\,WmK$^{-3}$. For the
four lowest temperatures, $T=0.37, 0.66, 0.96, 1.48$\,K, shown in
Fig.\ref{figExp}, we obtain from Eq.~(\ref{alpha}) $\alpha(T)\approx
0.52, 0.12, 0.049, 0.016$, respectively.

These estimates allow a quantitative comparision of theory and
experiment.  There are two main discrepancies between theory and
experiment which can be seen from a direct comparison of
Fig.~\ref{figExp} and Fig.~\ref{figScaling}.  First, there is a
 discrepancy in the position of the minimum (located at
$h\approx 3$ in the experiment and at $h\approx 4$ within the theory)
and second, the size of the dip of the order of 10\% is much smaller
than the predicted reduction of more than 50\% -- or the estimate for
$\alpha$ appears to be almost two orders of magnitude too small. What
can be the origin of the clear discrepancy?
First, one should note that for the temperatures and magnetic fields
shown in Fig.~\ref{figExp} both subleading effects  of order $\ln(B/T)/\ln(J/T)$
or $\ln(\ln( J/T))/\ln(J/T)$ and band-curvature effects (the overall
downturn
of $\kappa_{\rm mag}$ in large fields) neglected in our
calculation can become important.
For example, $\tilde{g}^2$ calculated to leading order is for
$J/T=30$ a factor 2.5 larger than the value obtained from Eq.~(\ref{g}).

More importantly, we believe that our model
(\ref{H0}) does not capture all aspects of the physics in the CuPzN samples
correctly. Especially, modeling the disorder by Eq.~(\ref{dis}) might not be appropriate.
This was also the conclusion of
Ref.~[\onlinecite{sologubenko}] from an analysis of the heat
conductivity at large fields $B\sim 15$\,T in the quantum critical
regime where the magnetization is close to saturation.
Indeed, for other types of disorder the matrix (\ref{mdis}) will have a
different structure which will affect the quantitative predictions while the qualitative picture will
remain unmodified. For example, it
might be necessary to take the interplay of forward scattering from impurities and interactions
into account. Forward scattering  affects transport at $B=0$ only very weakly but
can reduce the Umklapp dip in $\kappa_{\rm mag}$ considerably as the suppression
of $\kappa_{\rm mag}$ at larger fields relies on momentum conservation.
A more realistic modeling of disorder should also account for the possibility
 that defects cut the
one-dimensional chains in long pieces\cite{spin1}. In such a situation,  one has also to model how
phonons (or weak inter-chain interactions) couple heat into and out
off such long chain segments \cite{sologubenko,spin1}.

{\em Conclusions:} Our theoretical calculations show that the heat conductivity of weakly
disordered spin chains is very sensitive to moderate magnetic fields
$B \sim T$. A pronounced dip in the field dependence of $\kappa$ arises from the shift of the Fermi surface
of the spinons induced by the magnetic field. The effect of Umklapp scattering on the heat conductivity
turns out to be strongest when
the Fermi surface is shifted from the commensurate position at $B=0$ by an amount
of the order of its thermal broadening.

It is interesting to note that, according to our theory, this effect should be observable for a wide
range of parameters including spin-chains which have -- in contrast to
CuPzN -- large exchange couplings of several hundred Kelvin. Due to
the strong $B$ dependence, it should be possible to identify dips in
$\kappa_{\rm mag}$ even in the presence of a large phonon background. However,
for such systems, the effective disorder has to be sufficiently small,
$\alpha(T) \lesssim 10$. To obtain an effective disorder strength of the
order of 1 at $B \sim T$, typical fluctuations of the exchange coupling have to be
of the order of $B$ (for this crude estimate we used $D_{\rm dis} \sim
(\delta J)^2 a$ and neglected logarithmic renormalizations) $\delta
J/J \lesssim \mu_B B/J $. Furthermore, $\Delta k \sim \mu_B B/(J a)$
is also small for large $J$ and a necessary condition for the
quantitative validity of our calculations is that there is no
substantial forward scattering on the associate length scale $1/\Delta
k \sim a J/(\mu_B B)$. In systems with large $J$ and strong phonon
scattering,
one has also to take into account\cite{shimshoni03} that the sound
velocity $c$ is smaller than the spinon-velocity $v$. Therefore it may
 happen that the position of the dip in $\kappa_{\rm mag}$ moves to lower
 values, $h \sim c/v$, as the relevant energy scale \cite{shimshoni03} for phonon-assisted
 Umklapp scattering is $c \Delta k$  rather than $v \Delta k$.


\acknowledgements
We gratefully acknowledge discussions with N. Andrei and J. Sirker and financial support of the German--Israeli
Foundation and the DFG under SFB 608.


\begin{thebibliography}{99}

\bibitem{review}
I. Affleck, in {\it Fields, Strings and Critical Phenomena, Les
Houches, Session XLIX}, edited by E. Brezin and J. Zinn-Justin
(North-Holland, Amsterdam, 1988)

\bibitem{giamarchi} T. Giamarchi, {\it{Quantum Physics in one dimensional
systems}}, Oxford Univ. Press (2004).

\bibitem{bethe}
H. Bethe, Z. Phys. {\bf 71}, 205 (1931).

\bibitem{spin-ex}
A. V. Sologubenko, K. Gianno, H. R. Ott, U. Ammerahl and A.
Revcolevschi, Phys. Rev. Lett. {\bf 84}, 2714 (2000); K. Kudo, S.
Ishikawa, T. Noji, T. Adachi, Y. Koike, K. Maki, S. Tsuji and K.
Kumagai, J. Low Temp. Phys. {\bf 117}, 1689 (1999); A. V.
Sologubenko, E.Felder, K. Gianno, H. R. Ott, A. Vietkine and A.
Revcolevschi, Phys. Rev. B {\bf 62}, R6108 (2000); A. V.
Sologubenko, K. Gianno, H. R. Ott, A. Vietkine and A.
Revcolevschi, Phys. Rev. B {\bf 64}, 054412 (2001); C. Hess, C.
Baumann, U. Ammerahl, B. Büchner, F. Heidrich-Meisner, W. Brenig,
and A. Revcolevschi, Phys. Rev. B {\bf
  64}, 184305 (2001).


\bibitem{sologubenko}
A. V. Sologubenko,  K. Berggold, T. Lorenz, A. Rosch, E.
Shimshoni, M. D. Phillips and M. M. Turnb, Phys. Rev. Lett. {\bf
98}, 107201 (2007).

\bibitem{almost} P. Jung, A. Rosch,  Phys. Rev. B 76, 245108 (2007),
P. Jung, R. W. Helmes, A. Rosch, Phys. Rev. Lett.  96, 067202 (2006).

\bibitem{fisher} C. A. Doty and D. S. Fisher, Phys. Rev. B {\bf 45},
  2167 (1992).


\bibitem{zotos} A. V. Savin, G. P. Tsironis, and X. Zotos, Phys.Rev. B
  {\bf 75}, 214305 (2007).

\bibitem{ladder} E. Boulat, P. Mehta, N. Andrei, E. Shimshoni,
  A. Rosch,  Phys. Rev. B  {\bf 76}, 214411 (2007).


\bibitem{shimshoni03}
E.~Shimshoni, N.~Andrei, and A.~Rosch, Phys. Rev. B \textbf{68},
104401 (2003).

\bibitem{memory} R. Zwanzig,
J. Chem. Phys. {\bf 33}, 1338 (1960); H. Mori,
Prog. Theor. Phys. {\bf 33}, 423 (1965); D. Forster, {\it Hydrodynamic Fluctuations, Broken Symmetry,
and Correlation Functions}, (Benjamin, Massachusetts,
1975); W. G\"otze and P. W\"olfle, Phys. Rev. B {\bf 6}, 1226 (1972).


\bibitem{bounds} P. Jung, A. Rosch, Phys. Rev. B  75, 245104 (2007).


\bibitem{roschPRL} A. Rosch, N. Andrei, Phys. Rev. Lett. {\bf 85},
1092 (2000).

\bibitem{footnote} More precisely, $\kappa_{\rm mag}$ of the clean system
is finite but exponentially large if terms which break integrability and
higher-order Umklapp processes are taken into account \cite{roschPRL,shimshoni03}.

\bibitem{lukyanov}
S. Lukyanov,  Nucl. Phys. B {\bf 522}, 533 (1998).

\bibitem{spin1} 
A. V. Sologubenko, T. Lorenz, J. A. Mydosh, A. Rosch, K. C. Shortsleeves, M. M. Turnbull,
Phys. Rev. Lett. {\bf 100}, 137202 (2008).

\end{thebibliography}
\end{document}